\documentstyle[floats,prd,aps,epsf]{revtex}
\draft 

\begin{document}


\twocolumn[\hsize\textwidth\columnwidth\hsize\csname
@twocolumnfalse\endcsname

\title{Radiative Falloff in Neutron Star Spacetimes}

\author{Vasiliki Pavlidou$^1$, Konstantinos Tassis${}^1$,
	Thomas W.~Baumgarte$^2$, and Stuart L.~Shapiro$^{1,2,3}$}

\address{${}^1$ Department of Astronomy, 
	University of Illinois at Urbana-Champaign, Urbana, Il~61801}

\address{${}^2$ Department of Physics, University of Illinois at
	Urbana-Champaign, Urbana, Il~61801}

\address{${}^3$ NCSA, 
	University of Illinois at Urbana-Champaign, Urbana, Il~61801}

\maketitle

\begin{abstract}
We systematically study late-time tails of scalar waves propagating in
neutron star spacetimes.  We consider uniform density neutron stars,
for which the background spacetime is analytic and the compaction of
the star can be varied continously between the Newtonian limit $2M/R
\ll 1$ and the relativistic Buchdahl limit $2M/R = 8/9$.  We study the
reflection of a finite wave packet off neutron stars of different
compactions $2M/R$ and find that a Newtonian, an intermediate, and a
highly relativistic regime can be clearly distinguished.  In the
highly relativistic regime, the reflected signal is dominated by
quasi-periodic peaks, which originate from the wave packet bouncing
back and forth between the center of the star and the maximum of the
background curvature potential at $R \sim 3 M$.  Between these peaks,
the field decays according to a power-law. In the Buchdahl limit $2M/R
\rightarrow 8/9$ the light travel time between the center and the
maximum or the curvature potential grows without bound, so that the
first peak arrives only at infinitely late time.  The modes of neutron
stars can therefore no longer be excited in the ultra-relativistic
limit, and it is in this sense that the late-time radiative decay from
neutron stars looses all its features and gives rise to power-law
tails reminiscent of Schwarzschild black holes.  
\end{abstract}


\vskip2pc]


\section{Introduction}

The late-time radiative falloff in a black hole spacetime was first
studied with the goal of understanding how, in the absence of
rotation, a spherically symmetric Schwarzschild black hole emerges in
a general aspherical collapse.  The first numerical results by de la
Cruz, Chase and Israel~\cite{cci70} were followed up by the perturbative
analytic study by Price~\cite{p72}, who showed that scalar,
electromagnetic and gravitational radiation all decay as an inverse
power of time.  This late-time behavior is commonly refered to as
``power-law tails''.

The analysis of Gundlach, Price and Pullin~\cite{gpp94} showed that
power-law tails arise from backscattering off the weak gravitational
potential in the far zone ({\it cf.}~\cite{clsy95}).  One immediate
implication, namely that the late time behavior should be different in
black hole spacetimes which are not asymptotically flat, was confirmed
by Brady, Chambers, Krivan and Laguna~\cite{bckl97}, and later Brady,
Chambers, Laarakkers and Poisson~\cite{bclp99}.  These authors studied
radiative falloff in Schwarzschild-de Sitter spacetimes, and found
that for nonspherical (and minimally coupled) perturbations, the
power-law decay at early times changed into an exponential decay after
a cosmological horizon timescale~\cite{footnote1}.  Another
implication is that power-law tails should be present even in the
absence of event horizons, {\it e.g.}~for perturbations of neutron
stars.

Perturbations of nonrotating, spherically symmetric relativistic
neutron stars have been studied by many authors, mostly with the goal
of finding their oscillation
modes~\cite{tc67,d75,ld83,k88,cf91a,cf91b,ks92,lns93,akk96}.
Typically, these modes are determined by assuming a time-dependence
$\exp(i\sigma t)$ and then studying the time-independent problem.  The
perturbation equations then reduce to an eigenvalue problem, which can
be solved for the complex frequencies $\sigma$.  In addition to families
of modes corresponding to those of Newtonian stars, relativistic stars
also have modes which are reminiscent of the quasi-normal modes of
black holes and can be associated with the dynamical degrees of
freedom of the gravitational field.  Because of their association with
gravitational waves, these modes are usually refered to as $w$-modes.
Several distinct families of $w$-modes, both in axial and polar
parity, have been found in numerical calculations (see the recent
review by Kokkotas and Schmidt,~\cite{ks99}): curvature modes exist
for all compactions $2M/R$, where $M$ is the stellar mass and $R$ the
circumferential radius, trapped modes only arise in sufficiently
compact stars with $2M/R \gtrsim 2/3$, when radiation can be
``trapped'' inside the maximum of the background curvature potential
at $R \sim 3 M$ (see below), and interface modes are associated with a
discontinuity of the background curvature at the surface of the star.

Similar to black holes, the response of a neutron star to a general
perturbations will be a superposition of fluid and gravitational wave
modes and a power-law tail.  Since all modes decay exponentially, the
late-time falloff should be dominated by the power-law tail.  Unlike
the quasi-normal modes of a black hole, however, the frequency of the
neutron star's $w$-modes depends on the compaction $2M/R$ of the star.
As the compaction approaches the ultrarelativistic Buchdahl limit
$2M/R = 8/9$~\cite{b59}, the imaginary part of the frequency of at
least some of these modes approches zero, implying that their damping
time becomes infinite~\cite{akk96,ks99}.  A similar behavior had
earlier been identified by Detweiler for the fluid modes ($f$-modes)
of relativistic stars~\cite{d75}.  This observation immediately raises
the question how a neutron star in the ultrarelativistic limit
responds to a general perturbation, whether a power-law tail emerges,
and whether its characteristic radiative fall-off approaches that of a
black hole.  In this paper we seek to answer this question by
dynamically evolving the time-dependent perturbation equations.

So far, few authors have studied perturbations of relativistic stars
by dynamically evolving the time-de\-pen\-dent perturbation equations, as
opposed to assuming a $\exp(i\sigma t)$ time-dependence and solving
the time-independent equations.  Andersson and Kokkotas~\cite{ak96}
studied the excitation of $w$-modes by an impinging gravitational
wave, and discussed the prospect of detections by future
gravitational-wave laser interferometers.  Allen, Andersson, Kokkotas
and Schutz~\cite{aaks98} analyzed the spectrum of excited fluid and
wave modes arising from several different sets of initial data.  In
some of these simulations, both the quasi-normal modes and subsequent
power-law tails were observed (see~\cite{ak96}, also Fig.~5
in~\cite{ks99}).  Ruoff~\cite{r00} generalized earlier results and 
included realistic equations of state.

In this paper, we dynamically evolve a scalar field in a neutron star
background.  We model the neutron star as a homogeneous,
incompressible sphere, for which the background metric is known
analytically.  We derive an analytical generalized tortoise
coordinate, which brings the equations into a particularly simple
form.  We then vary the compaction of the star between the Newtonian
limit $2M/R \ll 1$ and the Buchdahl limit $2M/R = 8/9$.
Interestingly, the generalized tortoise coordinate distance between
the center and the surface of the star diverges in the Buchdahl limit,
implying that the light travel time becomes infinite.

We then study the reflection of finite wave packets off neutron stars
of various compactions, and identify three distinct regimes: In the
Newtonian regime $2M/R \ll 1$, we find a reflection of the initial
pulse from the center of the star, and a subsequent power-law falloff.
In an intermediate regime, so-called trapped modes of the neutron star
are excited and decay exponentially (see also~\cite{ak96,aaks98,r00}).
After these modes have decayed to sufficiently small values, the
late-time tail re-emerges as a power-law.  In this paper we focus on
the highly relativistic regime, in which the wave packet bounces back
and forth between the center of the star and the maximum of the
background curvature potential at $R \sim 3 M$, giving rise to
quasi-periodic peaks in the reflected signal.  Between these peaks,
the field again decays according to a power-law. In the Buchdahl limit
$2M/R \rightarrow 8/9$ the light travel time between the center and
the maximum or the curvature potential grows without bound, so that
the first peak arrives only at infinitely late time.  The modes of
neutron stars can therefore no longer be excited in the
ultra-relativistic limit, and it is in this sense that tails from
neutron stars loose all their features and give rise to power-law
tails.

The paper is organized as follows.  In Sec.~II we introduce the basic
equations, starting with a general spherically symmetric spacetime,
then specializing to homogeneous neutron stars, and describing our
numerical implementation of the equations.  We discuss our numerical
results and present characteristic examples for each of three
different compaction regimes in Sec.~III.  In Sec.~IV we briefly
summarize our findings.  We also include an appendix containing the 
derivation a conserved energy integral of the motion, which we use as a
numerical check.


\section{Basic Equations}

\subsection{Scalar Waves in Spherical Symmetry}

In spherical symmetry, the line element can be written 
\begin{equation} \label{metric}
ds^2 = - e^{2 \Phi} dt^2 + 
	e^{2 \Lambda} dr^2 + 
	r^2 (d\theta^2 + \sin^2 \theta d\phi^2).
\end{equation}
The wave equation for a massless and minimally coupled scalar field  
$\varphi$,
\begin{equation}\label{waveq}
\Box \varphi = \varphi^{;\alpha}_{~;\alpha} = 
\frac{1}{\sqrt{-g}} \left( \sqrt{-g}\, g^{\alpha \beta} 
\varphi_{,\alpha}\right)_{,\beta} = 0,
\end{equation}
can then be written
\begin{equation} \label{equ1}
- e^{- 2 \Phi} \tilde{\varphi}_{,tt} 
+ \frac{e^{-(\Phi + \Lambda)}}{r^2} 
\left( r^2 e^{\Phi - \Lambda} \tilde{\varphi}_{,r} \right)_{,r} 
- \frac{\ell(\ell+1)}{r^2} \tilde{\varphi} = 0,
\end{equation}
where we have decomposed the angular part of $\varphi$ into the
spherical harmonics
$\varphi(t,r,\theta,\phi)=\tilde{\varphi}(t,r)Y_{\ell
m}(\theta,\phi)$.

Introducing a ``generalized tortoise coordinate'' $r_*$ satisfying
\begin{equation} \label{equ2}
\frac{dr_*}{dr} = e^{\Lambda - \Phi},
\end{equation}
and substituting
\begin{equation} \label{defpsi}
\psi = r \tilde{\varphi}
\end{equation}
into eq.~(\ref{equ1}) yields
\begin{equation} \label{equ3}
- \psi_{,tt} + \psi_{,r_* r_*} = V_{\ell}(r)\, \psi,
\end{equation}
where the effective potential $V_{\ell}(r)$ is given by
\begin{equation}\label{genpot}
V_{\ell}(r) =  e^{2 \Phi} \left(\frac{\ell(\ell+1)}{r^2} + 
\frac{e^{-2\Lambda}}{r} (\Phi - \Lambda)_{,r} \right).
\end{equation}

\subsection{Uniform Density Neutron Stars}

We now specialize to a constant density neutron star, for which 
\begin{equation}
e^{\Phi} = \left\{ \begin{array}{ll}
	\displaystyle
	\frac{3}{2} \left( 1 - \frac{2M}{R} \right)^{1/2} 
	- \frac{1}{2} \left( 1 - \frac{2Mr^2}{R^3} \right)^{1/2} & r < R \\
	\displaystyle 
	\left( 1 - \frac{2M}{r} \right)^{1/2}  & r > R 
	\end{array} \right.
\end{equation}
and 
\begin{equation}
e^{\Lambda} = \left\{ \begin{array}{ll}
	\displaystyle
	\left( 1 - \frac{2Mr^2}{R^3} \right)^{-1/2} & r < R \\
	\displaystyle
	\left( 1 - \frac{2M}{r} \right)^{-1/2}  & r > R. 
	\end{array} \right.
\end{equation}
Here $R$ denotes the surface radius and $M$ denotes the 
gravitational mass of the star~\cite{mtw}.

\begin{figure*}[t]
\begin{center}
\leavevmode
\epsfxsize=5in
\epsffile{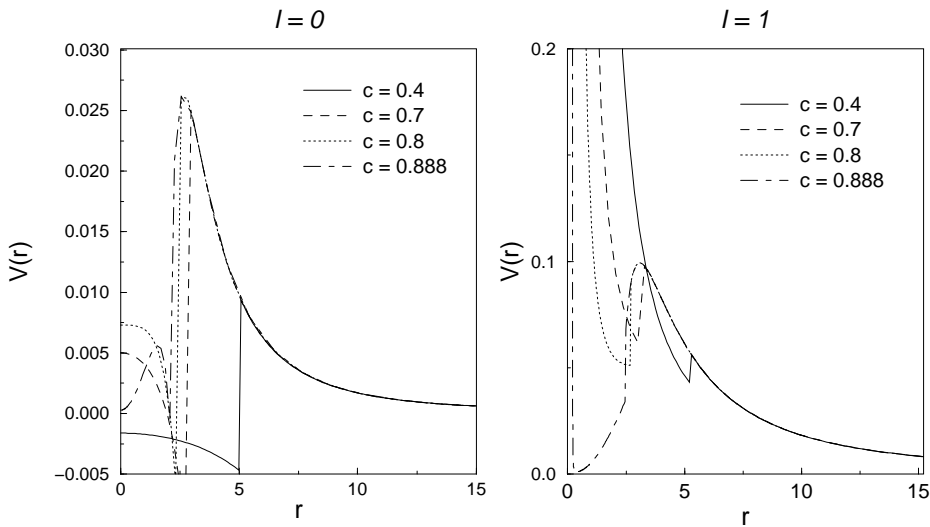}
\end{center}
\caption{The potential inside and outside of the star for different
values of the compaction $C = 2M/R$. Note the discontinuity at the
stellar surface.}
\end{figure*}

It is useful to introduce the nondimensional quantities 
\begin{equation}
\tilde{r}=\frac{r}{M}, \mbox{~~~~~~~}
\tilde{r}_{*}=\frac{r_{*}}{M}, \mbox{~~~~~~~}
\tilde{t}=\frac{t}{M}
\end{equation}
and to define the compaction of the star as
\begin{equation}
C=\frac{2M}{R}.
\end{equation}
Note that $C$ is limited by the well-known Buchdahl limit $C<8/9$~\cite{b59}.
With these definitions, the wave eq.~(\ref{equ3}) now becomes
\begin{equation} \label{siwaveqM}
- \psi_{,\tilde{t} \tilde{t}} + \psi_{,\tilde{r}_{*} \tilde{r}_{*}}
=\tilde{V}_{\ell}(r)\, \psi. 
\end{equation}
where we have also used $\tilde{V}_{\ell}(\tilde{r})=M^2 V_{\ell}(r)$.
In the following we will drop the tildes again, and it will be
understood that all variables are nondimensional.  This is equivalent
to setting $M \equiv 1$.

Adopting the notation of Chandrasekhar and Ferrari~\cite{cf91b},
\begin{equation}
y = \sqrt{1-C^3r^2/4}, \mbox{~~~~~~~}
y_1 = \sqrt{1-C},
\end{equation}
the generalized potential $V_{\ell}(r)$ can be written 
\begin{equation}\label{potential}
V_{\ell}(r) = \left\{ \begin{array}{lll}
        \displaystyle
        \frac{3y_1-y}{4} & \displaystyle\left(\frac{\ell(\ell+1)}
	{r^2}(3y_1-y)\right.\\
\displaystyle
&\left.\,\,\,+\displaystyle \frac{C^3}{4}(2y-3y_1)\right),
& r < 2/C
\\
        \displaystyle
  \lefteqn{\left( 1 - \frac{2}{r} \right) \left(
\frac{\ell (\ell 
+1)}{r^2} + \frac{2}{r^3} \right),} && r > 2/C.
        \end{array} \right.
\end{equation}
In the exterior of the star $r> 2/C$ the potential reduces to the
Schwarzschild potential.  Note also that the potential is discontinous
at the surface of the star ($r=2/C$).  The size of the discontinuity
is $3 C^3 (1-C)/8$, independent of $\ell$.  In Figure~(1) we show
$V_{\ell}(r)$ for different values of the compaction for $\ell=0$ and
$\ell=1$.

\subsection{The Generalized Tortoise Coordinate}

We now have to integrate eq.~(\ref{equ2}) to find an expression for 
the generalized tortoise coordinate $r_*$.  For $r>2/C$, the integration
\begin{equation}
r_* = \int e^{\Lambda - \Phi} dr = \int \frac{dr}{1 - 2/r} 
\end{equation}
yields the familiar Schwarzschild tortoise coordinate
\begin{equation}\label{rtortekso}
r_* = r + 2  \ln(\frac{r}{2} - 1)
\end{equation}
(where a constant of integration has been chosen appropriately).

In the interior, $r<2/C$, we have to integrate
\begin{equation}\label{toint}
r_* = \int \frac{2\,dr}{y \,(3 y_1 - y)}.
\end{equation}
With the substitution
\begin{equation}\label{thisist}
s = \frac{1}{3 y_1 - y},
\end{equation}
the integrand can be brought into the form
\begin{equation}
r_* = - \frac{4}{C^{3/2}}\int \frac{ds}{(-y_2 s^2 + 6 y_1 s - 1)^{1/2}},
\end{equation}
where we have used
\begin{equation}
y_2 = 8 - 9C.
\end{equation}
The integration can now be carried out analytically and yields
\begin{eqnarray}\label{rtort}
\nonumber r_{*} & = &  
\frac{4}{(C^3 y_2)^{1/2}} \left(
\arcsin \left(-\frac{y_2}{3 y_1 - y} + 3 y_1\right) \right. \\
& - & \left.
\arcsin \left(-\frac{y_2}{2 y_1}+ 3 y_1\right) \right)
 + \frac{2}{C}+2\ln(\frac{1}{C}-1),
\end{eqnarray}
where the dependence on $r$ enters through $y$, and where 
we have chosen a constant of integration such that $r_*$ is continous
across the surface of the star.  As expected, we recover $r_*/r
\rightarrow 1$ in the Newtonian limit $C \rightarrow 0$.  In the
ultrarelativistic limit $C\rightarrow 8/9$, $r_*$ at the origin $r=0$
approaches negative infinity, $r_*(0) \rightarrow - \infty$.  Since
$r_*$ measures the coordinate time which elapses along a photon radial
geodesic (eq.~(\ref{siwaveqM})), this implies that the light travel
time between the surface and the center of the star grows without
bound as the star's compaction reaches the Buchdahl limit.  This
property has important consequences for the scattering of waves from
ultrarelativistic stars, as we will discuss in Sec.~III.  In Fig.~2,
we show $r_*$ as a function of $r$ for various different compactions
$C$.

\begin{figure}
\begin{center}
\leavevmode
\epsfxsize=3in
\epsffile{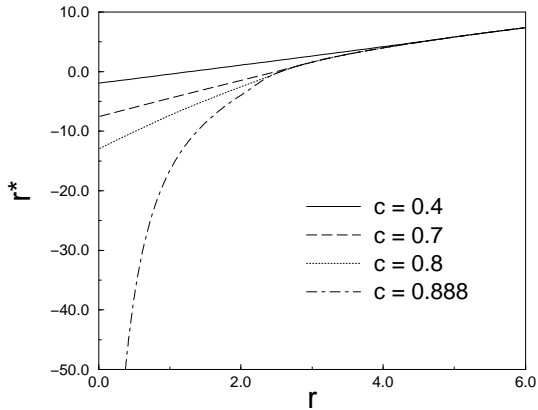}
\end{center} 
\caption{The tortoise coordinate r* as a function of r for several
different values of $C$.}
\end{figure}

\subsection{Numerical Implementation}

We integrate eq.~(\ref{siwaveqM}) numerically using the
finite-differencing scheme described by Gundlach, Price and
Pullin~\cite{gpp94}, which is based on the double 
null coordinates
\begin{equation}
u=t-r_{*},  \mbox{~~~~~~}
v=t+r_{*}.
\end{equation}
In terms of these, eq.~(\ref{siwaveqM}) becomes
\begin{equation}\label{nullwave}
-4\psi_{,uv}=V_{\ell}(r)\psi.
\end{equation}
Denoting
\begin{equation}
u_{+}=u+\Delta, \mbox{~~~~~~}
v_{+}=v+\Delta, 
\end{equation}
where 
$\Delta$ is the grid spacing, we can write a finite difference equation
for eq.~(\ref{nullwave}) as
\begin{eqnarray}\label{discret}
\nonumber
\psi(u_{+},v_{+}) & = & \displaystyle \left(1-\frac{\Delta^2}{8} V_{\ell}(r)
\right) \left(\psi(u_{+},v)+\psi(u,v_{+})\right) \\
& & - \psi(u,v).
\end{eqnarray}
Here $r$ can be found from $r_* = (v-u)/2$ by numerically inverting
eq.~(\ref{rtort}).  We impose a regularity boundary condition at the
origin ($\psi = 0$ at $r = 0$) and an outgoing wave boundary condition
\begin{equation}
\psi(u_+,v_+)= \psi(u_+,v)
\end{equation}
at a large radius $r = r_{\rm max}$ in the asymptotically flat regime.

\begin{figure*}[t]
\begin{center}
\leavevmode
\epsfxsize=5.5in
%
%
\epsffile{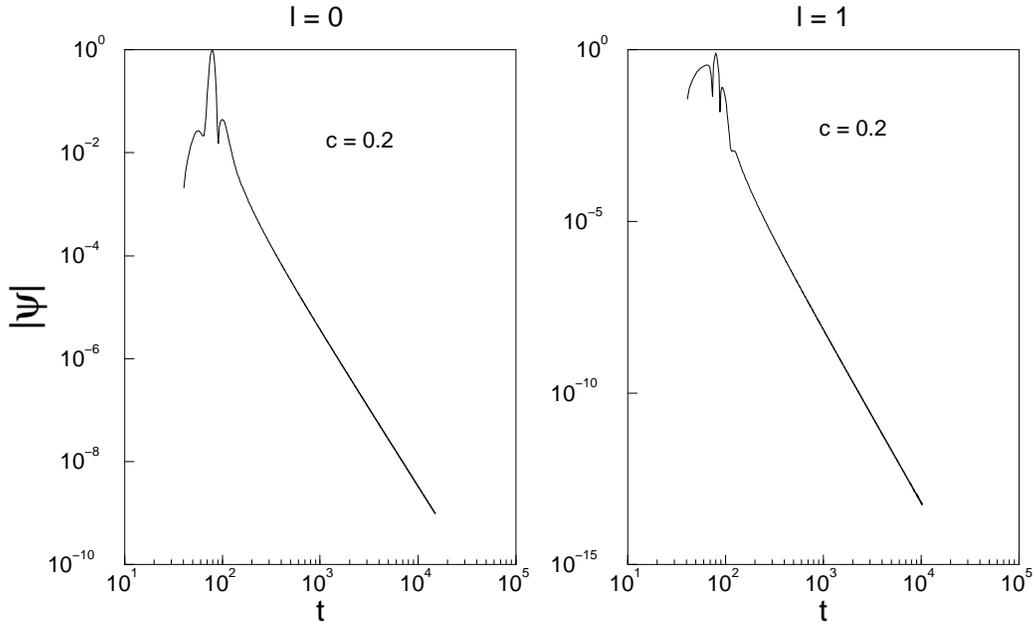}
\end{center}
\caption{
Late-time behaviour for the $\ell=0$ and $\ell=1$ modes 
with $C=0.2$. For such low values of the compaction we 
observe a power-law decay in a nearly Newtonian exterior
geometry.}
\end{figure*}

\begin{figure*}[t]
\begin{center}
\leavevmode
\epsfxsize=5.5in
%
%
\epsffile{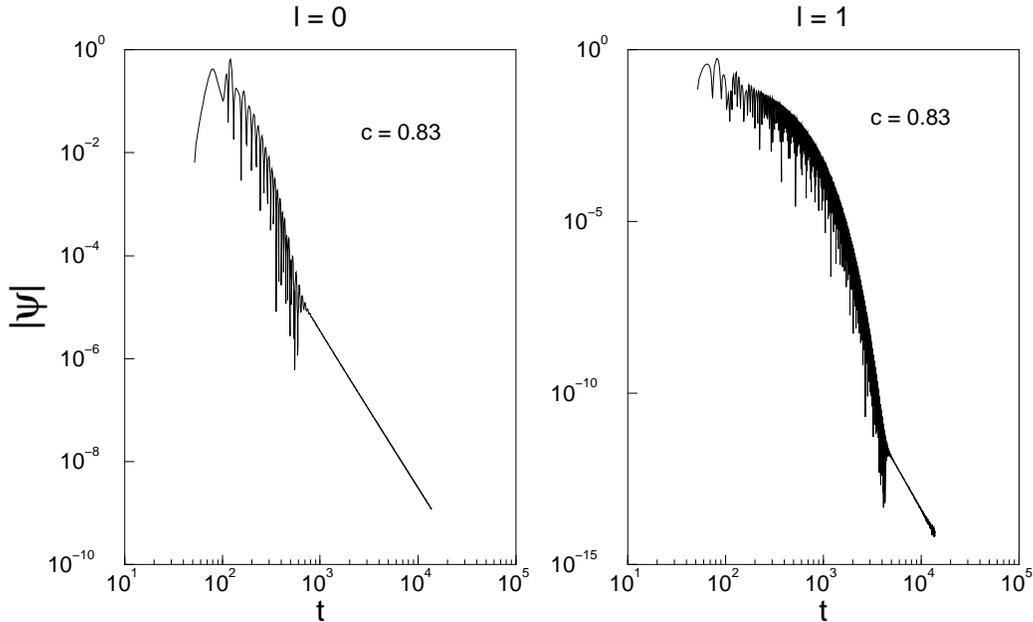}
\end{center}
\caption{
Late-time behaviour for the $\ell=0$ and $\ell=1$ modes
with $C=0.83$. Both modes are in the intermediate compaction
regime  where trapped quasi-normal modes are excited and are observed to 
decay exponentially.}
\end{figure*}

\begin{figure*}[t]
\begin{center}
\leavevmode
\epsfxsize=5.5in
%
%
\epsffile{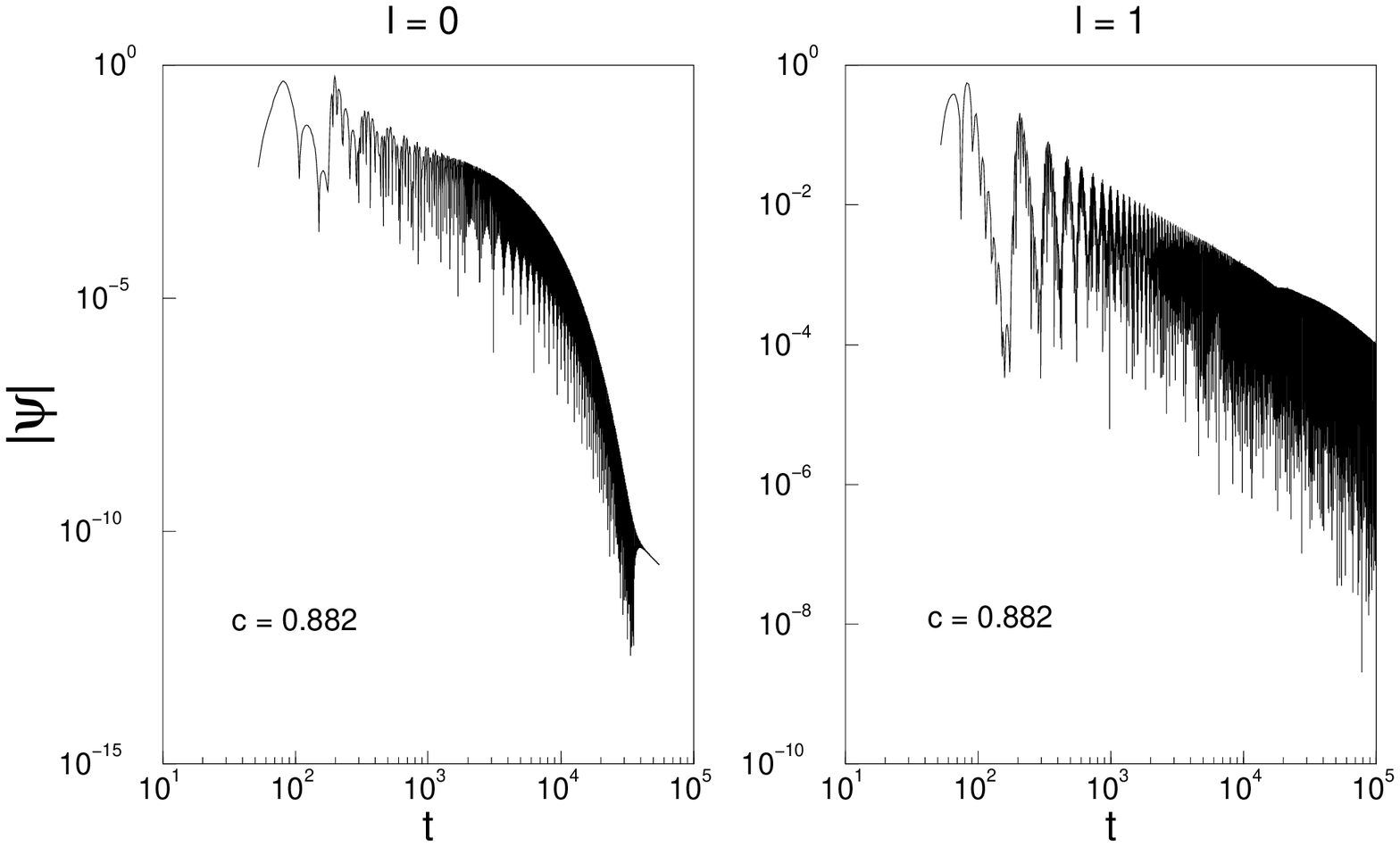}
\end{center}
\caption{Late-time behaviour for the $\ell=0$ and $\ell=1$ modes with
$C=0.882$. Both modes are in the strongly relativistic regime, 
where the early part of the decay is dominated by almost periodic peaks. 
In the case of the $\ell=0$ mode we can also see how the exponential decay
of quasi-normal modes is recovered in later times, to eventually give rise
to the power-law tail.}
\end{figure*}

We tested our code in various ways.  For a vanishing potential
$V_{\ell}(r)$ (flat space and $\ell = 0$), the analytical scalar 
wave solution is reproduced {\em exactly}.  For smooth potentials
(as for a Schwarzschild black hole), the code produces second
order accurate results ($O(\Delta^2)$).  This second order convergence
is broken by the discontinuity in the potential at the surface of 
the star, but the code is still convergent.  We also convinced ourselves
that by smoothing out the discontinuity, second order convergence
is restored.  Lastly, we checked that our evolution scheme conserves
energy, which is discussed in more detail in Appendix A.

Following~\cite{bclp99}, we choose as initial data $\psi(u,0)=0$ and
$\psi(0,v)= \exp\left(-(v-v_c)^2/(2\sigma^2) \right)$ for $v_c=30$ and
$\sigma=3$.  However, the qualitative features of our findings should
be independent of this particular choice.  All results will be shown
as a function of time $t=\frac{1}{2}(v+u)$ for an observer at fixed
radius $r_{*}=50$.


\section{Numerical Results}

\begin{table}
\begin{center}
\begin{tabular}{ccccc}
$2M/R$  &  $r_*(0)$  & $r_*(R)$ & $\alpha$ ($\ell=0$)
& $\alpha$ ($\ell=1$) 
\\
\tableline
 0.2 & 0.704 & 12.773 & -3.01 & -5.02 \\
 0.83 & -16.799 & -0.762 & -3.01 &  -5.02 \\
 0.882 & -55.947 & -1.755 & -3.00 & ---\\
 0.888 & -163.039 & -1.889 & -3.01 & ---\\
 0.8888 & -525.380 & -1.907 & -3.01 & --- 
\end{tabular}
\end{center}
\caption{Summary of different runs.  We tabulate the compaction $C=
2M/R$, the value of the tortoise coordinate at the origin $r_*(0)$ and
the surface of the star $r_*(R)$, and the slope $\alpha$ of the
late-time power law tail $t^{-\alpha}$ for $\ell =0$ and $\ell = 1$.
For $\ell=0$ with $C=0.888$ and $C=0.8888$, the slopes $\alpha$ refer
to the power law decay before the arrival of the first of the periodic
peaks.  The accuracy of the slopes is limited by the finite
integration time.}
\end{table}

We focus on the late-time radiative falloff of the scalar field, and 
find that three different regimes of  neutron star compaction which
can be clearly distinguished.  We discuss these different regimes below,
and summarize our numerical results in Table~1.

For small compactions we observe a power-law decay in a nearly
Newtonian (exterior) Schwarzschild geometry. The initial pulse is
reflected at the origin and then propagates outward, leaving behind
the power-law decay $\psi \sim t^{-\alpha}$ from backscattering off
the weak spacetime curvature at large radii.  The slope $\alpha$ of
this power-law is consistant with the value $-(2\ell+3)$ predicted for
backscatter in any spherically symmetric, static, asymptotically flat 
spacetime~\cite{gpp94}. As a typical example, we show
results for $C=0.2$ for both $\ell=0$ and $\ell=1$ in Fig.~3.  This
behaviour is observed for compactions $C$ up to about $0.8$ for the
$\ell=0$ mode and $0.7$ for the $\ell=1$ mode.

At these high compactions, we find a smooth transition into an
intermediate regime.  Once the star is sufficiently compact, a
sufficiently deep potential well forms between the peak of the
exterior Schwarzschild potential and the origin (see Fig.~1), and the
incoming pulse can excite quasi-normal modes trapped in this potential
well.  These modes decay exponentially as their energy slowly
``leaks'' out to infinity.  Once their amplitude has dropped to small
enough values, the power-law tail emerges and dominates the late-time
decay of the field.  In Fig.~4, we demonstrate this behavior for
$C=0.83$.

Note that due to the discontinuity of the potential, there is a small
potential well just inside the surface of the star for all compactions
$C$.  For small compactions, however, this well is too small to excite
quasi-normal modes, and it is only when the surface of the star is
close to the maximum of the exterior Schwarzschild potential that this
potential well becomes sufficiently large (compare Fig.~1) to observe
the exponential decay.

A third kind of behaviour is observed in the strongly relativistic
regime, when the compaction of the star is $C \gtrsim 0.86$ for $\ell=0$
and $C \gtrsim 0.83$ the $\ell=1$. Here the early part of the decay of the
initial pulse is dominated by quasiperiodic peaks with a period
approximately equal to twice the light travel time between the maximum
of the exterior effective potential and the origin.  Recall that for
increasingly large compactions, the origin is pushed to larger
negative values of $r_*$, thereby increasing the width of the
potential well.  For sufficiently large compactions, the wavelength of
the incoming pulse is smaller than the width of the potential well.
Instead of exciting a quasi-normal mode, it oscillates between the
origin, where it is completely reflected, and the maximum of the
external Schwazschild potential, where it is partially reflected.
Every time the pulse reaches the latter, a small part of it will be
transmitted as one of the quasi-periodic peaks which we observe at
large radii.  

During its travel in the non-zero potential background within the well
and with each reflection at the barrier, the pulse is distorted and
becomes broader with time.  This effect makes the observed pattern
quasi-periodic as opposed to periodic.  Once the typical wavelength
becomes comparable with the width of the well, we can no longer
cleanly separate individual peaks.  Simultaneously, the exponential
decay found in the intermediate regime is recovered, suggesting that
once the pulse ``fills out'' the well, the dynamics can again be
described by a exponentially decaying quasi-normal mode.  Ultimately,
a power-law tail again emerges for very late times.  We show this
behavior for $C=0.882$ in Fig.~5.  For $\ell=1$, the reappearence
of the power-law occurs only once the signal has decayed below
computer accuracy, which is why we are unable to observe this
transition.

An extreme case of the decay pattern described above is exhibited in
the case of extremely large compactions. In this case, as $C$
approaches $8/9$, the light travel time from the maximum of the
exterior potential to the origin grows without limit.  Thus, the time
between the initial peaks also grows without bound.  As a result, we
can observe a power-law decay between individual peaks.  When the next
peak arrives, it dominates over the power law tail until it decays
enough so that the power-law once again emerges. However, since the
power-law itself drops to increasingly small amplitudes, and since the
peaks broaden and eventually overlap, the power-law tail is
temporarily lost, to re-emerge only at very late times, when periodic
peaks and subsequent exponential decay have decayed sufficiently so
that the power law tail becomes once again the dominant pulse remnant.
The early part of this behaviour for $\ell=0$ and for compactions
$C=0.888$ and $C=0.8888$ is shown in Fig. 6.

In the limit $C \rightarrow 8/9$, the light travel time between
potential maximum and origin becomes infinite and hence the first
pulse will arrive infinitely late.  Thus, in the Buchdahl limit we
recover the characteristic power-law decay of a black hole spacetime.

\begin{figure}
\begin{center}
\leavevmode
\epsfxsize=3.2in
\epsffile{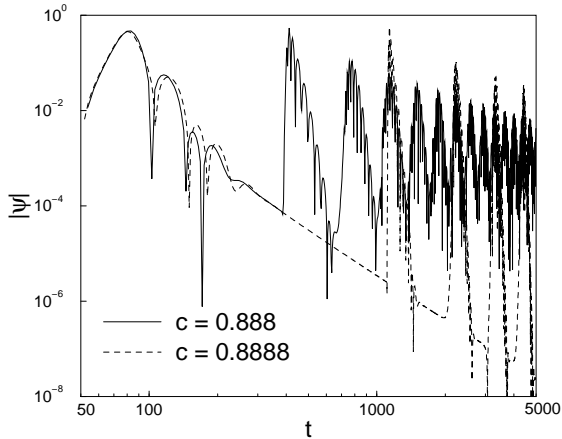}
\end{center}
\caption{Late-time behavior for $\ell = 0$ for stars with extremely
large compactions. The decaying field has enough time to enter the
power-law tail before the first of the quasiperiodic pulses
arrives. Note that as the compaction increases, the first pulse
arrives at increasingly late times. }
\end{figure}

\section{Summary}

We study the backscattering of scalar waves off uniform density
neutron stars, varying the compaction of the neutron star from the
Newtonian limit $2M/R \ll 1$ to the ultrarelativistic Buchdahl limit
$2M/R = 8/9$, and focussing on the excitation of quasi-normal modes
and the emergence of power-law tails.  We find that three distinct
regimes in the compaction can be identified.  

In the Newtonian regime, the response is dominated by a reflection of
the initial pulse off the center of the star, and subsequent power-law
falloff.  In an intermediate regime, more quasi-normal modes of the
neutron star excited, which decay exponentially and ultimately drop in
amplitude below the power-law tail, which dominates the late-time
falloff.  In the highly relativistic regime, in which the light travel
time between the center of the star and the maximum of the curvature
potential increases to large values, the initial wave package bounces
back and forth between those two and gives rise to quasi-periodic
signals.  Between these quasi-periodic signals, the field decays
following a power-law.  In the ultrarelativistic limit, the light
travel time reaches infinity, and the initial wave signal will never
reach the center of the star.  The modes of neutron stars can
therefore no longer be excited, and it is in this sense that the
late-time radiative decay from from ultrarelativistic neutron stars
looses all its features and again gives rise to power-law tails.  By
tracking the scattered wave amplitude, a distant observer can
distinguish whether the central object is a black hole or a neutron
star, and if the latter, can determine the compaction of the star.


\acknowledgments

The authors gratefully acknowledge useful conversations with Eric
Poisson.  This work was supported by NSF Grants AST 96-18524 and PHY
99-02833 and NASA Grant NAG 5-7152 at Illinois.


\begin{appendix}

\section{Conservation of Energy}

\begin{figure*}[t]
\begin{center}
\leavevmode
\epsfxsize=5.5in
\epsffile{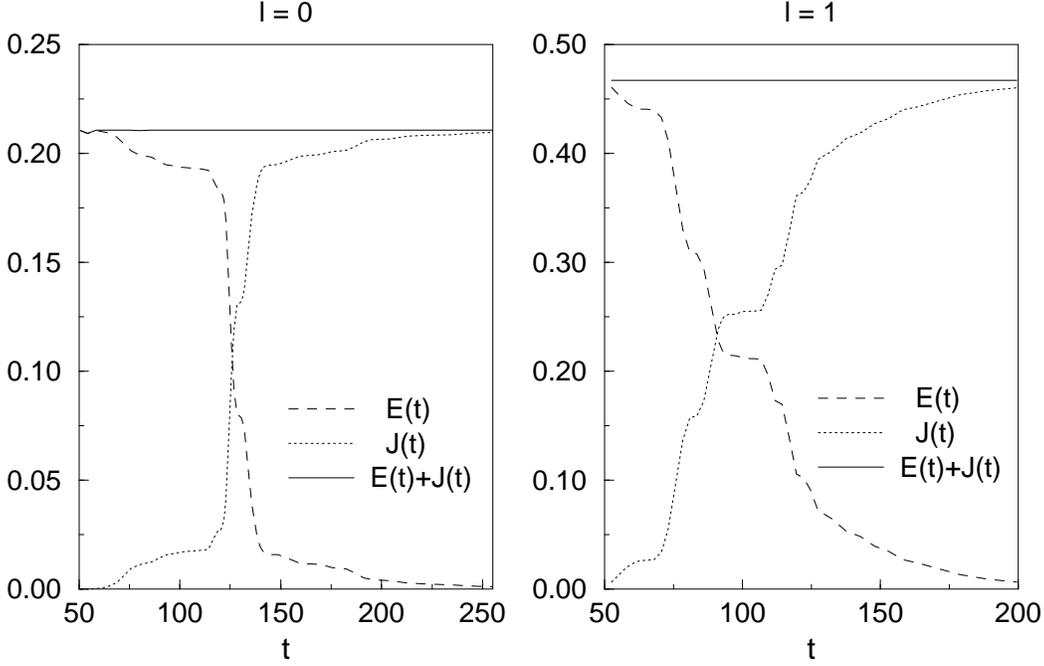}
\end{center}
\caption{Energy conservation for $\ell =0$, $C=0.85$ and for $\ell=1$,
$C=0.8$. We plot the total energy $E(t)$ contained within $r = 50$,
the integrated flux $J(t)$ across $r=50$, and the sum, which has to be
equal to $E(0)$, a constant.}
\end{figure*}

In this appendix we derive a conserved energy integral which can be
used as a test of the numerical simulation.

Since the spacetime~(\ref{metric}) is static, there exists a Killing
vector $\vec{\xi}=\partial/\partial t$. Thus, a conserved current can
be constructed by contracting $\vec{\xi}$ with the stress energy
tensor $T_{\mu\nu}$,
\begin{equation}\label{defJ}
J_{\mu}= T_{\mu\nu} \xi^{\nu}.
\end{equation} 
The conservation of $\vec{J}$ implies $ J^{\alpha}_{~;\alpha}=0$
or, by integrating over a 4-volume $V$,
\begin{equation}\label{conserv}
\int _V \left( g^{1/2}J^{\alpha} \right) \,_{,\alpha} d^4 x =0
\end{equation}
Applying Gauss's theorem, eq.~(\ref{conserv}) yields
\begin{equation} \label{gauss}
\oint_{\delta V} J^{\alpha} d^3 \Sigma_{\alpha}=0
\end{equation}
where $\delta V$ denotes the closed 3-surface enclosing the 4-volume $V$.  The
surface element is given by $d^3 \Sigma_{\alpha} = 1/3!
g^{1/2}\epsilon_{\alpha\beta\gamma\delta}dx^{\beta}dx^{\gamma}dx^{\delta}$,
where $\epsilon_{\alpha\beta\gamma\delta}$ is the Levi-Civita
alternating symbol.  In the following we choose $V$ to be the volume
defined by $0 \leq r \leq r_0$ and $t_0 \leq t \leq t_1$, so that
eq.~(\ref{gauss}) can be written as
\begin{eqnarray}\label{equ5}
\nonumber 
& \displaystyle\left[ \int_{r=0}^{r_{obs}} \int_{\theta=0}^{\pi}
\int_{\phi=0}^{2\pi} J^0 |g|^{1/2} d r d \theta d \phi 
\right]_{t_0}^{t_1}=&  \\
 & - \displaystyle\left. \int_{t_0}^{t_1} \int_{\theta=0}^{\pi}
\int_{\phi=0}^{2\pi}
J^r |g|^{1/2} d t  d \theta d \phi \right|_{r=r_{0}}. &
\end{eqnarray}
Here the left hand side represents the difference in energy interior
to a spherical volume between two instants of time $t_0$ and $t_1$,
and the right hand side represents the integrated flux through the
surface of the volume at $r = r_0$ during that time interval.

For a massless scalar wave the stress energy tensor is
\begin{equation}\label{timunu}
T_{\mu\nu}=\frac{1}{4\pi}[\varphi_{,\mu}\varphi_{,\nu}-\frac{1}{2}g_{\mu\nu}
\varphi_{,\alpha}\varphi^{,\alpha}],
\end{equation}
so that the relevant components of the flux $J^{\mu} = T^{\mu}_{~0}$ become
\begin{equation}\label{j0}
J^{0}=
-\frac{1}{8\pi}\left(e^{-2\Phi}\varphi_{,0}^{\,2}+
e^{-2\Lambda}\varphi_{,r}^{\,2}+
\frac{1}{r^2}\varphi_{,\theta}^{\,2}+
\frac{1}{r^2\sin ^2 \theta}\varphi_{,\phi}^{\,2}\right)
\end{equation}
and
\begin{equation}\label{jr}
J^r=\frac{1}{4\pi}e^{-2\Lambda}\Phi_{,r}\Phi_{,0}
\end{equation}
In the following we will consider an axisymmetric mode ($m=0$), for
which all derivatives with respect to $\phi$ vanish.

Inserting the fluxes together with the determinant of the metric
$|g|^{1/2}=e^{\Phi+\Lambda}r^2\sin \theta$ into eq.~(\ref{equ5}) and
performing the integration over $\phi$
then yields
\begin{eqnarray}\label{equ6}
\nonumber & & \displaystyle \left[ \frac{1}{2} \int
dr d\mu r^2 \left( e^{\Lambda-\Phi}\varphi_{,0}^{2}
+e^{\Phi-\Lambda}\varphi_{,r}^{2}
+\frac{e^{\Phi+\Lambda}}{r^2}\varphi_{,\theta}^{2} \right)
\right]_{t_0}^{t_1} \\
& & = \displaystyle 
\left. \int dt d\mu r^2 
\varphi_{,r}\varphi_{,0}
e^{\Phi-\Lambda}\right|_{r=r_{0}}
\end{eqnarray}
where we have used $\mu=\cos\theta$.  The limits of integration in
eq.~(\ref{equ6}) are from $r=0$ to $r_0$, from $t=t_0$ to $t_1$ 
and from $\mu=-1$ to $1$.

We can now define  $E(t)$ as the energy inclosed within a sphere of 
radius $r_0$ at time $t$,
\begin{equation} \label{energy}
E(t)=\frac{1}{2}\left.\int
dr d\mu r^2 \left( e^{\Lambda-\Phi}\varphi_{,0}^{\,2}
+e^{\Phi-\Lambda}\varphi_{,r}^{2}
+\frac{e^{\Phi+\Lambda}}{r^2}\varphi_{,\theta}^{\,2}\right)
\right|_{t} \end{equation}
(with the same limits of integration as in eq. \ref{equ6}), and $J(t)$
as the outgoing flux of energy through the surface of the same sphere
integrated over the surface of this sphere and over the time
interval $t-t_0$,
\begin{equation}\label{flux}
\left.
J(t)=-\int_{t_0}^{t}dtd\mu\,
r^2\varphi_{,r}\varphi_{,0}e^{\Phi-\Lambda}\right|_{r=r_0}.
\end{equation}
Eq.~(\ref{equ6}) can now be rewritten
\begin{equation}\label{ejsum}
E(t)+J(t)=E(t_0)=const
\end{equation}

The expression for the energy and flux integrals can be further simplified
be decomposing $\varphi$ into spherical harmonics
\begin{equation}
\varphi(t,r,\theta,\phi)=\tilde{\varphi} (t,r)Y_{\ell m}(\theta,\phi)
\end{equation} 
as we do in our dynamical evolution.

For the spherically symmetric case $\ell=0$, $Y_{00} = 1/\sqrt{4 \pi}$,
the energy integral~(\ref{energy}) becomes 
\begin{equation}\label{energyl0}
E_{0}(t)=\frac{1}{4\pi}\int_{0}^{r_{0}}
r^2 \left(e^{\Lambda-\Phi}\tilde{\varphi}_{,0}^{\,2}
+ e^{\Phi -\Lambda} \tilde{\varphi}_{,r}^{\,2} \right) dr 
\end{equation}
and the flux integral~(\ref{flux})
\begin{equation}\label{fluxl0}
\left. J_{0}(t)=-\frac{1}{2 \pi} \int_{t_0}^{t} r^2
\tilde{\varphi}_{,r}\tilde{\varphi}_{,0}e^{\Phi-\Lambda} \,
dt \right|_{r=r_0}
\end{equation}

For $\ell=1$ and $m=0$ we have $Y_{10}= \sqrt{3/(4\pi)}\mu$, and
eqs.~(\ref{energy}) and~(\ref{flux}) now become
\begin{eqnarray}\label{energyl1}
\nonumber  E_{1}(t)= \frac{1}{4\pi}\int_{0}^{r_{0}}
r^2 \left( e^{\Lambda-\Phi}\tilde{\varphi}_{,0}^{2}
+e^{\Phi - \Lambda} \tilde{\varphi}_{,r}^{2}\right)dr + \\
+\frac{1}{2\pi}\int_{0}^{r_{0}}  e^{\Phi + \Lambda}\tilde{\varphi}^{2} dr
\end{eqnarray}
and
\begin{equation} \label{fluxl1}
J_{1}(t)=\frac{1}{2\pi}\left.\int_{t_0}^{t}\, r^2
\tilde{\varphi}_{,r}
\tilde{\varphi}_{,0} e^{\Phi-\Lambda}dt\right|_{r=r_0}
\end{equation}

Energy conservation in our dynamical code can now by checked by
verifying that eq.~(\ref{ejsum}) is satisfied at all times.  In
Fig.~(7) we show the energy and flux integrals for examples with
$\ell=0$ and $\ell=1$.  The sum of the two integrals remains constant
with an accuracy which increases with increasing grid resolution,
implying that energy is indeed conserved.

\end{appendix}


\end{document}